\documentstyle[aip]{article}

\newtheorem{prop}{Proposition}
\newtheorem{teo}{Theorem}
\newcommand{\ep}{\epsilon}
\newcommand{\lab}{\label}

\newcommand{\be}{\begin{equation}}
\newcommand{\ee}{\end{equation}}
\newcommand{\bea}{\begin{eqnarray}}
\newcommand{\eea}{\end{eqnarray}}

\def\Journal#1#2#3#4{{#1} {\bf #2}, #3 (#4)}

\def\JMP{\em J. Math. Phys.}

\def\PLB{{\em Phys. Lett.}  B}
\def\ra{\rightarrow}

\def\lab{\label}
\begin{document}
%\tableofcontents % INDICE
\setcounter{page}{0}

%%%%%%%%%%%%%%%%%%%%%%%%%%%%%%%%%%%%%%%%%%%%%%%%%%%%%%%%%%%%%%%%%
\title{\Large\bf Properties of equations of the continuous Toda type}
\hspace{.5cm}
\author{\large $^{*}$E. Alfinito, $^{\dag}$ G. Profilo, $^{*}$ G. Soliani\\
{\it $^{*}$  Dipartimento di Fisica dell'Universit\`a, 73100 Lecce,
Italy,}\\
{\it and Istituto Nazionale di Fisica Nucleare, Sezione di Lecce,
Italy.}\\
{ $^{\dag}$ \it  Dipartimento di Matematica dell'Universit\`a,
73100 Lecce, Italy.}}

\maketitle

%{\centering PACS \vskip3pc}

\begin{abstract}
\small{We study a modified version of an equation of the continuous
Toda type in 1+1 dimensions. This equation contains a friction-like
term which can be
switched off by annihilating a free parameter $\ep$. We apply the
prolongation method, the symmetry and the approximate symmetry approach.
This strategy allows us to get insight into both the equations for $\ep =0$
and $\ep \ne 0$, whose properties arising in the above frameworks are
mutually compared. For $\ep =0$, the related prolongation equations are
solved by means of certain series expansions which lead to an infinite-
dimensional Lie algebra. Furthermore, using a realization of the Lie
algebra of the Euclidean group $E_{2}$, a connection is shown between the
continuous Toda equation and a linear wave equation which resembles
a special case of a three-dimensional
wave equation that occurs in a generalized Gibbons-Hawking ansatz
\cite{lebrun}.
Nontrivial solutions to the wave equation expressed in terms
of Bessel functions are determined.

For $\ep\,\ne\,0,$ we obtain a finite-dimensional Lie algebra
with four elements. A
matrix representation of this algebra yields solutions of the modified
continuous Toda equation associated with a reduced form of a perturbative
Liouville equation. This result coincides with that achieved in the context
of the approximate symmetry approach. Example of exact solutions are also
provided. In particular, the inverse of the exponential-integral function
turns out to be defined by the reduced differential equation coming from a
linear combination of the time and space translations. Finally, a Lie
algebra characterizing the approximate symmetries is discussed.}
\end{abstract}

\section{INTRODUCTION}

We investigate the equation
\be u_{tt}\,+\,\ep u_{t}\,=\,(e^{u})_{xx}
\lab{1}\ee
where $u\,=\,u(x,t)$, subscripts denote partial derivatives and $\ep$ is a
constant. For $\ep\,=\,0$, Eq.(\ref{1}) is a continuous Toda system in 1+1
dimensions (or, equivalently, a two-dimensional version of
the so-called heavenly equation: self-dual Einstein spaces with a rotational
Killing vector \cite{Saveliev}). The latter arises in many branches of
physics, running from the theory of Hamiltonian systems, to the topological
field theory \cite{Krichever}. In the case in which $\ep \ne 0$, the term
$\ep u_{t}$ mimics a friction-like behaviour. Eq.(\ref{1}), for $\ep\ne 0$
has been handled, in part, in Ref.[3] by means of
an approximate group analysis.

We look for an algebraic characterization of  Eq. (\ref{1}) in both the
cases $\ep\,=\,0$ and $\ep\,\ne\,0$. In doing so, we resort to the
prolongation method \cite{Estabrook} and the symmetry approach
\cite{Ibragimov,Olver}. For simplicity, we shall keep the formal machinery
inherent to these techniques to a minimum. Our main results are the
following. For $\ep\,=\,0$, Eq. (\ref{1}) can be written as a set of
(prolongation) differential equations which can be solved in terms of
power series
expansions whose coefficients (in the variable $z=e^{u}$) depend on the
pseudopotential components and obey a presumably
infinite-dimensional Lie algebra.

A remarkable fact is that this algebra can be closed "step by step", in the
sense that a finite-dimensional Lie algebra emerges in correspondence of
each polynomial in $z$ coming from the truncation of the series. The use of
a given representation of any closed Lie algebra allows us to find a
linear problem associated with the equation under consideration.
Furthermore, we show that the prolongation differential equations related to
Eq. (\ref{1}) ($\ep=0$) afford a class of solutions connected with the Lie
algebra of the Euclidean group $E_{2}$ in the plane. This enables us to map
Eq. (\ref{1}) into the linear wave equation
\be y_{tt}\,-\,e^{u} y_{xx}\,=\,0,
\lab{2}\ee
where $y\,=\,y(x,t)$ is a pseudopotential variable. We point out that
Eq.(2) is equivalent to a 1+1 dimensional version of Eq.(2) of ref.[1] in
which a generalized Gibbons-Hawking ansatz \cite{gibbons} pertinent to a
quantum theory of gravity is considered. Starting from simple
solutions of Eq. (\ref{1}), examples of nontrivial solutions of Eq. (\ref{2})
are displayed.

In the case $\ep \ne 0$, the prolongation algebra for Eq. (\ref{1}) turns
out to be finite-dimensional. This algebra is exploited to determine a
class of special solutions {\it via} a reduced form of a Liouville-type
equation. The latter coincides with that arising in the context of the
symmetry approach. This feature suggests the existence of a possible link
between the prolongation method and the symmetry approach, which deserves
further studies. Finally, a self-similar solution in terms of the
exponential-integral function is obtained.

The paper is organized as follows. In Sec. II, the prolongation equations
derived for Eq. (\ref{1}) are studied. In Sections III and IV, the cases
$\ep=0$ and $\ep \ne 0$ are considered, respectively. Section V, deals with
the symmetry and the approximate symmetry approach applied to Eq. (\ref{1}).
Precisely, the generators of
the Lie point symmetries and approximate symmetries are found. The latter
can be characterized by a finite dimensional Lie algebra which admits a
realization in terms of boson annihilation and creation operators. In Sec.
VI, the main results are summarized and some comments are reported.
Finally,
Appendices A and B contain details of calculation, while Appendix C is
devoted to a brief introduction to the approximate symmetry method.

\section{THE PROLONGATION EQUATIONS}

Let us consider the prolongation system for Eq. (\ref{1}):
\be y^{i}_{x}\,=\,F^{i}(u,u_{t};y^{j}), \quad
 y^{i}_{t}\,=\,G^{i}(u,u_{x};y^{j}),
\lab{3}\ee
where $i,j= 1,2,\ldots, N$ ($N$ arbitrary), and the set of variables
$\{y^{i}\}$ is the pseudopotential \cite{Estabrook}. The compatibility
condition for Eq. (\ref{3}) yields
\be F^{i}\,=\,L^{i} u_{t}\,+\,M^{i},\qquad
G^{i}\,=\,L^{i} e^{u} u_{x}\,+\,P^{i},
\lab{4}\ee
where $M^{i}\,=\,M^{i}(u;y^{j})$, $P^{i}\,=\,P^{i}(u;y^{j})$, and
 $L^{i}\,=\,L^{i}(y^{j})$ are defined by
\be
M^{i}_{u}\,+\,[P,L]^{i}\,=\,\ep\,L^{i},\label{5}\ee
\be e^{u}\,[L,M]^{i}\,=\,P^{i}_{u},\lab{6}\ee
\be [M,P]^{i}\,=\,0,\lab{7}\ee
where $[P,L]^{i}\,=\,P^{k}\,\frac{\partial L^{i}}{\partial y_{k}}\,-\,
L^{k}\,\frac{\partial P^{i}}{\partial y_{k}}$, and so on.

For brevity, hereafter we shall omit the index $i$. Now, we look for a
solution of Eq. (\ref{5}-\ref{7}) of the form
\be M\,=\,\sum^{\infty}_{k=0}\,a_{k}(y)z^{k}, \qquad
P\,=\,\sum^{\infty}_{k=0}\,b_{k}(y)z^{k},
\label{8}\ee
where $z=e^{u}$ and $y$ stands for the set of components $\{y^{j}\}\;
(j\,=\,1,2,\ldots,N).$

By inserting the expansions (\ref{8}) into Eqs.(\ref{5}-\ref{7}), we obtain
the following constraints between the coefficients $a_{k}(y)$ and
$b_{k}(y)$:
\be [b_{0},L]\,=\,\ep L,\lab{9}\ee
\be [L,b_{k}]\,=\,k a_{k},\lab{10}\ee
\be [L,a_{k-1}]\,=\,k b_{k},\lab{11}\ee
\be [a_{0},b_{0}]\,=\,0,\lab{12}\ee
\be [a_{0},b_{1}]\,+\,[a_{1},b_{0}]\,=\,0,\lab{13}\ee
\be [a_{0},b_{2}]\,+\,[a_{1},b_{1}]\,+\,[a_{2},b_{0}]\,=\,0,\lab{14}\ee
$$ \ldots\ldots\ldots\ldots\ldots\ldots\ldots\ldots\ldots\ldots\ldots$$
\be \sum^{N}_{k=1}[a_{k-1},b_{N-k}]\,=\,0,\lab{15}\ee
where $k\,=\,1,2,\ldots N$ ($N$ arbitrary).

In order to scrutinize the commutation relations (\ref{9}-\ref{15}), two cases
have to be distinguished, i.e. $\ep=0$ and $\ep\ne 0$.

\section{THE CASE $\ep=0$}

Let us assume $\ep\,=\,0$ in Eqs. (\ref{9})-(\ref{15}).
Then, the systematic application of the Jacobi identity to the commutation
relations (\ref{9})-(\ref{15}) produces an arbitrary number of finite-
dimensional Lie algebras with 2N+1 generators (N=1,2,\ldots), i.e. $a_{0},
\,a_{1},\ldots,a_{N},\;b_{0},\;b_{1}\ldots,b_{N}$, and $L$ ($N$ arbitrary).

Although at present a rigorous proof is not given, this statement can be
checked heuristically "step by step" in the sense explained below.

To this aim, let us take (first step) $a_{j}\,=\,0,\;b_{j}\,=\,0,$ for
$j\,=\,1,2.\ldots$. Thus, from Eqs. (\ref{9})- (\ref{15}) we get the
abelian Lie algebra
\be [a_{0}, b_{0}]\,=\,0,\quad [a_{0}, L]\,=\,0,\quad [b_{0}, L]\,=\,0.
\lab{151}\ee

Any realization of this algebra corresponds to a solution of the type
$M\,=\,a_{0}(y)$, $P\,=\,b_{0}(y)$, to Eqs.(\ref{5})-(\ref{7}) (with
$\ep\,=\,0$). In this case, the prolongation system (\ref{3}-\ref{4})
becomes
\be y_{x}\,=\,L u_{t}\,+\,a_{0}, \quad y_{t}\,=\,L e^{u} u_{x}\,+\,b_{0}.
\lab{152}\ee
Of course, the compatibility condition for the system (\ref{152}) is
fulfilled if (\ref{151}) holds.

Now, let us choose (second step) $a_{j}\,=\,0,\;b_{j}\,=\,0,$ for $j\,=\,2,
3,\ldots.$ From (\ref{9})-(\ref{15}) we have (see Appendix A)
$$ [a_{0},a_{1}]\,=\,0,\quad [a_{0},b_{0}]\,=\,0, \quad [a_{0},b_{1}]\,=\,0,
\quad [a_{0},L]\,=\,- b_{1}$$
$$ [a_{1},b_{0}]\,=\,0,\quad [a_{1},b_{1}]\,=\,0, \quad [a_{1},L]\,=\,0,
\quad [b_{0},b_{1}]\,=\, 0$$
\be [b_{0},L]\,=\,0,\quad [b_{1},L]\,=\,-a_{1}. \lab{153}\ee
Equations (\ref{153})
represent a non-abelian Lie algebra defined by the five elements $a_{0},\,
a_{1},\,b_{0},\,b_{1}$ and $L$. Any realization of the algebra (\ref{153})
corresponds to a solution of the type
\be M(u,y)\,=\,a_{0}(y)\,+\,a_{1}(y)e^{u},
\lab{154}\ee
\be P(u,y)\,=\,b_{0}(y)\,+\,b_{1}(y)e^{u},
\lab{155}\ee
to Eqs.(\ref{5})-(\ref{7}). The prolongation equations (\ref{3})-(\ref{4})
related to (\ref{154})-(\ref{155}) are
\be y_{x}\,=\, L u_{t}\,+\,a_{0}\,+\,a_{1}e^{u},\qquad
y_{t}\,=\, L e^{u} u_{x}\,+\,b_{0}\,+\,b_{1}e^{u}.
\lab{156}\ee
The compatibility condition for this system is verified by the algebra
(\ref{153}).

Furthermore, as it is shown in Appendix A, under the hypothesis
$a_{j}\,=\,0,\quad b_{j}\,=\,0,$ for $j\,=\,3,4,\ldots$ (third step),
Eqs.(\ref{9})-(\ref{14}) give rise to the prolongation Lie algebra defined
by the seven elements $a_{0},\,a_{1},a_{2},\,b_{0},\,b_{1},\,b_{2},$ and
$L$:
$$ [a_{0},a_{1}]\,=\,0,\quad [a_{0},a_{2}]\,=\,0, \quad [a_{0},b_{0}]\,=\,0,
\quad [a_{0},b_{1}]\,=\,0$$
$$ [a_{0},b_{2}]\,=\,0,\quad [a_{0},L]\,=\,-b_{1}, \quad [a_{1},a_{2}]\,=\,0,
\quad [a_{1},b_{0}]\,=\, 0$$
$$ [a_{1},b_{1}]\,=\,0,\quad [a_{1},b_{2}]\,=\,0, \quad [a_{1},L]\,=\,-
2b_{2},
\quad [a_{2},b_{0}]\,=\, 0$$
$$[a_{2},b_{1}]\,=\,0,\quad [a_{2},b_{2}]\,=\,0,\quad [a_{2},L]\,=\,0,
\quad [b_{0},b_{1}]\,=\,0, $$
$$[b_{0},b_{2}]\,=\,0,\quad [b_{0},L]\,=\,0,\quad [b_{1},b_{2}]\,=\,0, $$
\be[b_{1},L]\,=\,a_{1},\quad [b_{2},L]\,=\,-2 a_{2}.\lab{157}\ee
The prolongation equations (\ref{3})-(\ref{4}) can be written as
\be y_{x}\,=\,L u_{t}\,+\,a_{0}\,+\,a_{1}e^{u}\,+\,a_{2}e^{2u},\quad
 y_{t}\,=\,L e^{u} u_{x}\,+\,b_{0}\,+\,b_{1}e^{u}\,+\,b_{2}e^{2u},
\lab{158}\ee
whose compatibility condition is ensured by (\ref{157}).

The next step consists in finding a closed Lie algebra defined by the nine
generators $a_{0},\, a_{1},\,a_{2},\,a_{3},\,b_{0},\,b_{1},\,b_{2},\,b_{3},$
and $L$ (see Appendix A). Since the character of the commutation relations
(\ref{b1})-(A 12) is basically of the recursive type, we expect that the
procedure of
building up finite-dimensional Lie algebras starting from (\ref{9})-(\ref{15})
($\ep$ =0) works out for any step. In other words we have the following
possible scenario: Equation (\ref{1}), ($\ep=0$) admits the prolongation
Lie algebra
defined by the commutation relations (see Appendix A):
$$[L,a_{l-1}]\,=\,l b_{l},\; [L,a_{N}]\,=\,0,\;[L,b_{k}]\,=\,k a_{k}, $$
\be [a_{j},a_{k}]\,=\,0,\; [a_{j}, b_{k}]\,=\,0,\;[b_{j},b_{k}]\,=\,0,
\label{159}\ee
for $j,k,= 0,1,2,\ldots,N,$ and $l= 1,2,\ldots,N$. Since $N$ is an
arbitrary positive integer, Eqs.(\ref{159}) represent an
infinite-dimensional Lie algebra.

\hspace{.3cm}

Another interesting result concerning the case $\ep\,=\,0$, is expressed by
the following

\begin{prop}
Let $u$ be a solution of the equation
\be u_{tt}\,=\,(e^{u})_{xx}.
\lab{16}\ee
Then, the function $y_{2}=y_{2}(x,t)$ defined by
\be y_{2x}\,=\,-i {\cal C}_{0}(\xi)\cosh y_{1},\lab{17}\ee
\be y_{2t}\,=\,-\frac{1}{2}\xi {\cal C}_{1}(\xi)\sinh y_{1},\lab{18}\ee
\be y_{1x}\,=\, u_{t}, \qquad y_{1t}\,=\,e^{u} u_{x},\lab{19}\ee
satisfies the wave equation
\be y_{2tt}\,-\,e^{u} y_{2xx}\,=\,0,\lab{20}\ee
where ${\cal C}_{0}(\xi)$, with $\xi\,=\,2ie^{\frac{u}{2}}$, fulfills the
differential equation of the Bessel type
\be \frac{d^{2} {\cal C}_{0}}{d \xi^{2}}\,+\,\frac{1}{\xi}\frac{d
{\cal C}_{0}}{d \xi}\,+\,{\cal C}_{0}\,=\,0,
\lab{21}\ee
and
\be {\cal C}_{1}\,=\,i \frac{d{\cal C}_{0}}{d \xi}.
\lab{22}\ee
\end{prop}

To prove this Proposition, let us search a solution to Eqs.(\ref{5}-\ref{7})
($\ep=0$) of the form
\be M(u,y)\,=\,m(u)g(y), \qquad P(u,y)\,=\,p(u)h(y). \label{23}\ee
Substitution from (\ref{23}) into Eqs. (\ref{5}-\ref{7}) yields
\be m_{u}\,=\,p, \qquad m e^{u}\,=\,p_{u}, \label{24}\ee
and
\be [L,h]\,=\,g,\quad [L,g]\,=\,h,\quad [g,h]\,=\,0. \label{25}\ee
Equations (\ref{24}) provide
\be m_{uu}\,-\,m e^{u}\,=\,0, \qquad p_{uu}\,-\,p_{u}\,=\,e^{u} p,
\label{26}\ee
which give
\be m\,=\,{\cal C}_{0}(\xi), \qquad p\,=\,-\frac{i}{2} \xi {\cal C}_{1}(\xi),
\label{27}\ee
respectively, where $\xi\,=\,2i e^{\frac{u}{2}}$, and $\cal C$ denotes a
function of the Bessel type $J,\, Y,\,H^{(1)},\,H^{(2)}$, or any linear
combination of them \cite{Abramowitz}.

We notice that by setting $L\,=\,i X_{1}$,  $h\,=\, X_{2}$ and $g\,=\,i
X_{3}$, Eqs.(\ref{25}) become:
\be [X_{1},X_{2}]\,=\,X_{3},\quad [X_{1},X_{3}]\,=\,X_{2} ,\quad
[X_{2},X_{3}]\,=\,0, \label{28}\ee
i.e. the Lie algebra associated with the Euclidean group, $E_{2}$, in the
plane \cite{Wybourne}.

A realization of Eqs. (\ref{28}) in terms of a two-component
pseudopotential $y\,=\,(y_{1}, y_{2})$ is
\be X_{1}\,=\,-i\partial_{y_{1}}, \quad X_{2}\,=\,-i \sinh y_{1}
\partial_{y_{2}}, \quad X_{3}\,=\,-\cosh y_{1}\partial_{y_{2}}.
\lab{29}\ee
 Then, Eqs.(\ref{3}) can be written as [see (\ref{4}) and (\ref{3})]:
\be y_{1x}\,=\,u_{t}, \quad y_{1t}\,=\,e^{u} u_{x},\label{30}\ee
\be y_{2x}\,=\,-i m \cosh y_{1}, \quad y_{2t}\,=\,-ip \sinh y_{1}.
\label{31}\ee

Thus, Eqs. (\ref{17}-\ref{18}) are nothing but (\ref{31}) with $m$ and $p$
replaced
by the quantities (\ref{27}). Furthermore, Eq. (\ref{20}) arises
straightforwardly form Eqs. (\ref{31}) with the help of Eqs. (\ref{24}).
Finally, Eqs. (\ref{21}) and (\ref{22}) are a direct consequence of
Eqs. (\ref{4}) and (\ref{9})-(\ref{15}).

We observe also that by combining together Eqs.(\ref{20}) and (\ref{16})
we can express  Eqs.(\ref{16}) by means of the pseudopotential component
$y_{2}$,
namely,
\be \partial_{t}^{2}\, \ln\frac{y_{2tt}}{y_{2xx}}\,=\,
\partial_{x}^{2}\, \frac{y_{2tt}}{y_{2xx}}.
\lab{32}\ee

Below we shall display a few examples of nontrivial solutions of the wave
equation (\ref{20}), starting from some simple solutions of Eq.(\ref{16}).

To this aim, let us consider the simple solution $u=t$ to Eq.(\ref{16}).
Then, by choosing ${\cal C}_{0}(\xi)\,=\,i J_{0}(\xi)$, so that
${\cal C}_{1}(\xi)\,=\,-\frac{d J_{0}(\xi)}{d \xi}\,=\,J_{1}(\xi)$
[see Eq.(\ref{27})], where $J_{0}$ and $J_{1}$ are Bessel
functions of the first kind \cite{Abramowitz}, from  Eq.(\ref{31}) we
obtain
\be y_{2x}\,=\,J_{0}(\xi) \cosh x, \lab{33}\ee
\be y_{2t}\,=\,- \frac{\xi}{2} J_{1}(\xi) \sinh x, \lab{34}\ee
where $\xi\,=\,2i e^{\frac{t}{2}}$ [$y_{1x}\,=\,1,\;\;y_{1t}\,=\,0;$ see
Eq. (\ref{30})].

Equations (\ref{33})-(\ref{34}) can be easily integrated to give
\be y_{2}(x,t)\,=\,J_{0}(2i e^{\frac{t}{2}}) \sinh x. \lab{35}\ee
Hence, the pseudopotential variable (\ref{35}) represents a particular
solution to the wave equation
\be y_{2tt}\,-\,e^{t} y_{2xx}\,=\,0. \lab{36}\ee

Another application of the previous Proposition concerns the equation
\be y_{2tt}\,-\,x y_{2xx}\,=\,0, \lab{37}\ee
which corresponds to $u=\ln x$ [a special solution to Eq.(\ref{16})]. Using
the same procedure as before, after easy calculations we find
\be y_{2}\,=\,\frac{1}{2}\xi\frac{d J_{0}(\xi)}{d \xi} \cosh t,
\lab{38}\ee
where $\xi\,=\,2i x^{\frac{1}{2}}$.

At this point it is instructive to show that the explicit form of the
pseudopotential can be used to solve certain linear second order ordinary
differential equations with variable coefficients. In fact, by way of
example, let us put
\be y_{2}\,=\,f(x) g(t) \label{39}\ee
into Eq. (\ref{37}). Then, Eq. (\ref{37}) entails
\be g_{tt}\,-\,g\,=\,0,\lab{40}\ee
\be f_{xx}\,-\,\frac{1}{x}f\,=\,0.\lab{41}\ee

Now, since $g\,=\,\cosh t$ is a particular integral of Eq.(\ref{39}), from
(\ref{38}) and (\ref{39}) we get
\be f(x)\,=\,-i x^{\frac{1}{2}} J_{1}(2ix^{\frac{1}{2}}). \lab{42}\ee

\section{THE CASE $\ep\ne0$}

For $\ep\ne 0$, the prolongation algebra of Eq.(\ref{1}) is a Lie algebra
$\cal L$ closed at the beginning (see Appendix A). This reads
\be
[a_{0},b_{0}]\,=\,[a_{0},b_{1}]\,=\,[L,b_{1}]\,=\,0,\label{43}\ee
\be [b_{0},b_{1}]\,=\,\ep b_{1}, \label{44}\ee
\be [b_{0},L]\,=\,\ep L ,\label{45}\ee
\be [L,a_{0}]\,=\, b_{1}.\label{46}\ee
A matrix representation of $\cal L$ is
$$ a_{0}\,=\,\left(\begin{array}{ccc} 0& -1&0\\ 0&0&0\\0&0&0\end{array}
\right),\quad
b_{0}\,=\,\left(\begin{array}{ccc} \ep& 0&0\\ 0&\ep&0\\0&0&0\end{array}
\right),$$
\be b_{1}\,=\,\left(\begin{array}{ccc} 0& 0&1\\ 0&0&0\\0&0&0\end{array}
\right),\quad
L\,=\,\left(\begin{array}{ccc} 0& 0&0\\ 0&0&1\\0&0&0\end{array}
\right).\lab{47}\ee
In view of (\ref{47}), Eqs.(\ref{3}) take the form
\be \left(\begin{array}{l} y_{1}\\ y_{2}\\y_{3}\end{array}
\right)_{x}\,=\,
\left(\begin{array}{ccc} 0& 0&0\\ -1&0&0\\0&u_{t}&0\end{array}
\right)\left(\begin{array}{l} y_{1}\\ y_{2}\\y_{3}\end{array}
\right),
\lab{48}\ee
\be \left(\begin{array}{l} y_{1}\\ y_{2}\\y_{3}\end{array}
\right)_{t}\,=\,
\left(\begin{array}{ccc} \ep& 0&0\\ 0&\ep&0\\e^{u}&e^{u}u_{x}&0\end{array}
\right)\left(\begin{array}{l} y_{1}\\ y_{2}\\y_{3}\end{array}
\right),
\lab{49}\ee
from which $y_{1}\,=\,\lambda_{1}\, e^{\ep t}$,  $y_{2}\,=\,\zeta\, e^{\ep t}$
and
\be y_{3\zeta}\,=\,-\frac{1}{\lambda_{1}}\, e^{\ep t}\, u_{t}\, \zeta,
\lab{50}\ee
\be y_{3 t}\,=\,-\lambda_{1}\, e^{\ep t}\, \zeta^{2}
\frac{\partial}{\partial \zeta} \frac{e^{u}}{\zeta},
\lab{51}\ee
where $\zeta\,=\,-\lambda_{1} x\,+\,\lambda_{2}$, and $\lambda_{1},
\lambda_{2}$ are constants of integration. Here $y_{3}$ plays the role of
a potential variable.

We remark that Eqs.(\ref{50})-(\ref{51}) can be exploited to determine
explicit solutions of Eq.(\ref{1}) ($\ep \ne 0$). In doing so, let us seek,
for instance, solutions of the type $u_{t}\,=\,\gamma(t)$. After some
manipulations, Eqs. (\ref{50}) and (\ref{51}) provide
\be
u\,=\,\ln\left[\frac{1}{2\lambda_{1}^{2}} (\gamma_{t}\,+\,\ep \gamma)
\zeta^{2}\,+\,\alpha \zeta\,+\,e^{-\ep t} \Gamma_{t}\right],
\lab{52}\ee
where $\gamma =\gamma(t)$, $\alpha=\alpha(t)= e^{\beta}$ and
$\Gamma=\Gamma(t)$ are
such that
\be\gamma=\frac{\alpha_{t}}{\alpha},\lab{53}\ee
\be \beta_{tt}\,+\,\ep \beta_{t}\,=\,c_{1} e^{\beta},
\lab{54}\ee
\be \Gamma_{t}\,=\,c_{2}\, e^{\int\,dt \, (\gamma\,+\,\ep)},
\lab{55}\ee
with $c_{1}$, $c_{2}$ constants of integration.

It is noteworthy that Eq. (\ref{54}), which is a modified version of the
reduced Liouville equation, has the same form of that obtained in the
framework of the Lie group approach \cite{Ibragimov} {\em via } a certain
symmetry variable (see Sec.V).

In the simple case $c_{1}\,=\,0$, Eq.(\ref{54}) is easily solved. Then we
find the exact solution
\be u\,=\,k_{1} e^{-\ep t}\,+\,\ln \left(e^{\frac{k_{2}}{\ep}}\,\zeta\,+\,
k_{0}\right)
\lab{56}\ee
to Eq.(\ref{1}), where $k_{0},\,k_{1}$ and $k_{2}$ are arbitrary constants.
 Otherwise (when $c_{1}\,\ne 0$ and $\ep$ is a small parameter),
Eq.(\ref{54}) can be analyzed by using some perturbative technique.

\section{THE SYMMETRY APPROACH}
\subsection{Symmetry generators}

As is well known \cite{}, in the study of a system of partial
differential equations one can use symmetry groups to find special
solutions (which are invariants under some subgroups of the complete
symmetry group) by solving reduced systems of differential equations
involving fewer independent variables than the original system. However,
the standard technique has to be modified if small perturbations are
present in the equations under consideration. In this context, the authors
of Ref. [2] devised a method based on the concepts of approximate group
of transformations and approximate symmetries. For the reader's convenience,
in Appendix C we shall recall the main aspects of this method.

In order to obtain the approximate symmetries admitted by Eq.(\ref{1})
$(\ep\ne 0)$, let us take in Eqs. (\ref{c20}) and (\ref{c21}):
%%%%%%%%%%%%%%%%%%%%%%%%%%%%%%%%%%%%%%%%%%%%%%%%%%%%%%%%%%%%%%%%
$$F_{0} \,=\,u_{tt}\,-\,(e^{u})_{xx},\qquad F_{1}\,=\,u_{tt}$$
$$z\,=\,(t,x,u,u_{x},u_{tt},u_{tx},u_{xx}),\qquad\qquad k\,=\,1,2,\ldots,8,
$$
\begin{equation}
X\,=\,(\xi^{1}_{0}\,+\,\epsilon\,\xi^{1}_{1})\partial_{t}\,+\,
(\xi^{2}_{0}\,+\,\epsilon\,\xi^{2}_{1})\partial_{x}\,+\,
(\xi^{3}_{0}\,+\,\epsilon\,\xi^{3}_{1})\partial_{u},\label{57}
\end{equation}
\begin{equation}
\xi^{j}\,=\,(\xi^{3}_{o})^{j}\,+\,\epsilon(\xi^{3}_{1})^{j}, \qquad\qquad
j\,=\,4,\ldots,8,\label{58}
\end{equation}
where
\begin{equation}
(\xi^{3}_{\alpha})^{J}\,=\,D_{J}\left(\xi^{3}_{\alpha}\,-\,\sum^{2}_{i=1}\,
u^{\alpha}_{i}\right)\,+\,\sum^{2}_{i=1}\,\xi^{i}_{\alpha}\,
u^{\alpha}_{J,i},\label{59}
\end{equation}
$$u^{\alpha}_{i}\,=\,\frac{\partial u^{\alpha}}{\partial x^{i}},\qquad
u^{\alpha}_{J,i}\,=\,\frac{\partial u^{\alpha}_{J}}{\partial x^{i}},\qquad
\alpha\,=\,0,1, $$
\noindent $D_{J}$ denotes the total derivative with respect to
$J\,=\,t,x,tt,tx,xx$, respectively, and $x^{1}\equiv t$, $x^{2}\equiv x$.

\noindent By expliciting (\ref{c20}) and (\ref{c21}) and keeping in
mind (\ref{57})-(\ref{59}), equating the coefficients of the
independent variables $u_{0t},\;u_{0t}u_{0x},$... to zero, we arrive at the
following set of constraints:
$$\xi^{3}_{j}\,+\,2\xi^{1}_{jt}\,-\,2\xi^{2}_{jx}\,=\,0, \qquad
\xi^{2}_{jxx}\,-\,2\xi^{2}_{jx}\,=\,0, $$
\begin{equation}
2\xi^{3}_{jtu}\,-\,\xi^{1}_{jtt}\,=\,0,\qquad
\xi^{3}_{jtt}\,-\,e^{u_{0}}\xi^{3}_{jxx}\,=\,0,\qquad\qquad (j=0,1),
\label{60}\end{equation}
and
\begin{equation}
2\xi^{3}_{jtu}\,-\,\xi^{1}_{jtt}\,+\,c_{j}=\,0,\label{61}
\end{equation}
where
$\xi^{1}_{j}=\xi^{1}_{j}(t)$, $\xi^{2}_{j}=\xi^{2}_{j}(x)$,
$\xi^{3}_{j}=\xi^{3}_{j}(x,t)$, and $c_{j}$ is a constant of integration.

\noindent By solving Eqs. (\ref{60}) and (\ref{61}), the
expression (\ref{57}) takes the form
\begin{eqnarray}
X\,& =& \,\left[c_{1}t\,+\,c_{2}\,+\,\epsilon(\frac{c_{1}}{2}t^{2}\,+\,k_{1}t\,
+\,k_{2})\right]\partial_{t}\,\nonumber\\
&+&\,
\left[(c_{1}\,+\,c_{3})x\,+\,c_{4}\,+\,\epsilon(\frac{k_{3}}{2}\,+\,k_{1})x\,
+\,\epsilon k_{4}\right]\partial_{x} \nonumber\\
& +& \,
\left[2c_{3}\,+\,\epsilon(-2c_{1}t\,+\,k_{3})\right]\partial_{u},
\label{62}\end{eqnarray}
with $c_{1},\,c_{2},\ldots\; {\rm and}\;k_{1},\,k_{2},\ldots$
arbitrary constants.

{}From the quantity (\ref{62}) we have the operators
 \bea X_{1}\,& =&\,X^{0}_{1}\,+\,\epsilon(\frac{t^{2}}{2}\partial_{t}\,-\,2t
\partial_{u}),\qquad
X_{2}\,\equiv\,X^{0}_{2}\,=\,\partial_{t} \nonumber\\
X_{3}\,&\equiv&\,X^{0}_{3}\,=\,\partial_{x} \qquad
X_{4}\,\equiv\,X^{0}_{4}\,=\,x\partial_{x}\,+\,2\partial_{u}
\label{63}\eea
where
\begin{equation}
\,X^{0}_{1}\,=\,t\partial_{t}\,+\,x\partial_{x}\label{64}\end{equation}
and
\bea
X_{5}\,&=&\,\epsilon(t\partial_{t}\,+\,x\partial_{x}),\qquad
X_{6}\,=\,\epsilon\partial_{t},\nonumber \\
X_{7}\,&=&\,\frac{\epsilon}{2}(x\partial_{x}\,+\,2\partial_{u}),\qquad
X_{8}\,=\,\epsilon\partial_{u}.\label{65}\eea
The operators $X^{0}_{1}$, $X^{0}_{2}$, $X^{0}_{3}$, $X^{0}_{4}$ are the
(exact) symmetry generators relative to Eq.(\ref{1}) for $\epsilon\,=\,0$
, while $X_{1}$, $X_{2}$, $X_{3}$, $X_{4}$ are the {\it approximate}
generators
of Eq.(\ref{1}) for $\epsilon\,\ne\,0$. The {\it exact} symmetry
generators of Eq. (\ref{1}) ) for $\epsilon\,\ne\,0$ are $X_{1}^{0},
\;X_{2}^{0}$
and $X_{3}^{0}$. The operators
(\ref{65}) are inessential, in the sense that $\epsilon$ is a constant factor.

\subsection{ Algebraic properties}
   The operators (\ref{63}), (\ref{64}) obey the commutation relations
\begin{equation}
[X_{2},X_{1}]\,=\,X_{2}\,+\,\epsilon(t\partial_{t}\,-\,2\partial_{u}),
\label{66}\end{equation}
\begin{equation}
[X_{2},X_{4}]\,=\,0,\quad [X_{2},X_{3}]\,=\,0 \quad [X_{1},X_{4}]\,=\,0,
\label{67}\end{equation}
\begin{equation}
[X_{1},X_{3}]\,=\,- X_{3},\qquad [X_{3},X_{4}]\,=\,X_{3}.
\label{68}\end{equation}
We notice that the commutation rules (\ref{66})-(\ref{68}) do not define a
(finite) Lie algebra.
However, they can be used to build up a realization of a finite-dimensional Lie
algebra by introducing the "auxiliary" operators
\begin{equation}
Z\,=\,t\partial_{t}\,-\,2\partial_{u}, \qquad Y\,=\,\frac{t^{2}}{2}
\partial_{t}\,-\,2t\partial_{u}.\label{69}
\end{equation}
In doing so, it turns out that
$Y$, $Z$, $X_{j}\;
(j=1,\dots,4)$ satisfy the commutation relations
\be
[X_{2},X_{1}]\,=\, X_{2}\,+\,\epsilon Z,\quad [X_{3},X_{1}]\,=\, X_{3},
\quad [X_{1},X_{4}]\,=\, 0, \lab{70}\ee
\begin{equation}
[X_{1},Y]\,=\, Y,\quad [X_{1},Z]\,=\,- \epsilon Y, \quad
\label{71}\end{equation}
\be[X_{2},Y]\,=\, Z,\quad [X_{2},Y]\,=\, X_{2},\quad [Z,Y]\,=\, Y,\lab{72}
\ee
$$
[X_{2},X_{3}] \,=\,\quad [X_{2},X_{4}]\,=\,[X_{3},X_{4}]\,=\,[X_{3},Y]\,
=\, 0,$$
\be [X_{4},Y]=\,[X_{3},Z]\,=\,[X_{4},Z]\,=\,0. \label{73}\ee

For brevity, by the symbols $Y$, $Z$, $X_{j}\;(j=1,\dots,4)$ we
shall indicate both the abstract elements and the corresponding
realization (\ref{63})-(\ref{64}) and
(\ref{66})-(\ref{68}) of the finite-dimensional Lie
algebra (\ref{70})-( \ref{73}).

\noindent Now, let us focus our attention on the subalgebra
(\ref{72}). This is isomorphic to the $sl(2,R)$ algebra given by
\begin{equation}
[Z^{\prime}, T]\,=\,2S, \quad [T,S]\,=\,2Z^{\prime},\quad
[S,Z^{\prime}]\,=\,-2T, \label{74}
\end{equation}
where
\begin{equation}
 T\,=\,\sqrt{2}(Y\,+\,X_{2}),\quad S\,=\,\sqrt{2}(Y\,-\,X_{2}),\quad
Z^{\prime}\,=\,2Z. \label{75}
\end{equation}

Furthermore, the following Propositions hold:

\begin{prop}

The Casimir operator
\begin{eqnarray}
C\,&=&\, T^{2}\,-\,S^{2}\,-\,{Z^{\prime}}^{2}\nonumber\\
&\equiv&\,4[2X_{2}Y\,-\,Z(Z\,+\,1)]\end{eqnarray}
\begin{equation}
\label{76}\end{equation}
of the Lie algebra (\ref{74}), commutes with {\it all} the  generators
$Y$, $Z$, $X_{j}\;(j=1,\dots,4)$ of the Lie algebra
 (\ref{70})-(\ref{73}).
\end{prop}

The proof is straightforward.

\begin{prop}

The algebra (\ref{70})-(\ref{73}) admits a realization in terms of
boson annihilation and creation operators.
\end{prop}

This can be shown by putting
\begin{equation}
a^{\dag}_{1}\,=\,t,\;\;a^{\dag}_{2}\,=\,u,\;\;a^{\dag}_{3}\,=\,x,\;\;
a_{1}\,=\,\partial_{t},\;\;a_{2}\,=\,\partial_{u},\;\;
a_{3}\,=\,\partial_{x},\;\;
\label{77}\end{equation}
to yield
\begin{equation}
[a_{j},a^{\dag}_{k}]\,=\,\delta_{jk},\quad
[a_{j},a_{k}]\,=\,0\quad\quad (j,k=1,2,3),
\label{78}\end{equation}
$$Y\,=\,\frac{1}{2}{a^{\dag}}^{2}_{1}a_{1}\,-\,2a^{\dag}_{1}a_{2},\qquad
Z\,=\,a^{\dag}_{1}a_{1}\,-\,2a_{2},$$
$$X_{1}\,=\,a^{\dag}_{1}a_{1}\,+\,a^{\dag}_{3}a_{3}\,+\,\epsilon\,
\left(\frac{1}{2}{a^{\dag}}^{2}_{1}a_{1}\,-\,2a^{\dag}_{1}a_{2}
\right),$$
\be X_{2}\,=\,a_{1},\quad X_{3}\,=\,a_{3},\quad X_{4}\,=\,
a^{\dag}_{3}a_{3}\,+\,2a_{2},
\label{79}\end{equation}
and
\begin{equation}
C\,=\,-8a_{2}(a_{2}\,+\,1).\label{80}
\end{equation}
We would like to finish this Subsection with a remark on the "auxiliary"
operators (\ref{69}). These are essential for establishing the closed
algebra (\ref{70})-(\ref{73}). Notwithstanding, their meaning is not
completely clear. For instance it should be important to ascertain whether
finite-dimensional Lie algebra analogous to (\ref{70})-(\ref{73}) can be
constructed in relation to other case studies. At present, we are able to
affirm only that the operators (\ref{69}), $Z$ and $Y$ can be interpreted,
respectively, as symmetry variables of the equations
\be u_{xx}\,+\,\ep u_{x}\,=\,-(e^{-u})_{tt},
\lab{811}\ee
\be u_{xx}\,+\,\ep u_{x}\,=\,-2(e^{-\frac{u}{2}})_{tt},
\lab{822}\ee
which arise, formally from Eq. (\ref{1}) {\em via} the transformations
$t\, \ra\,x$, $u \,\ra\,-u$, and $t\, \ra\,x$, $u \,\ra\,-\frac{u}{2}$.
{}From $Z$ we obtain the invariant $\eta(x)\,\equiv\,e^{q(x)}\,=\,t
e^{\frac{u}{2}} $, that once replaced into Eq.(\ref{811}) yields the reduced
equation of the modified Liouville type
\be q_{xx}\,+\,\ep\,q_{x}\,=\,-e^{-2q}.
\lab{833}\ee
In a similar way, from (\ref{822}) we find
\be r_{xx}\,+\,\ep\,r_{x}\,=\,-4e^{-\frac{r}{2}},
\lab{844}\ee
where the invariant $r(x)\,=\,u\,+\,4 \ln t$ associated with $Y$ has been
used.
\subsection{Explicit solutions}
   To see how the symmetry approach works out, we shall deal with two
significant examples. First, let us write down the group transformations
related to the generator $X_{4}\equiv X^{0}_{4}$ [see (\ref{63})], which is
present in both the cases $\ep=0$ and $\ep\ne 0$. We have
\be x^{\prime}\,=\,e^{\lambda} x,\;t^{\prime}\,=\, t,\;
u^{\prime}\,=\,u\,+\,2\lambda,
\lab{81}\ee
where $\lambda$ is the group parameter.

{}From (\ref{81}) we deduce the invariant
\be \rho\,=\,x^{\prime}\,=\,e^{-\frac{u^{\prime}}{2}} \,=\,
x\,=\,e^{-\frac{u}{2}}.
\lab{82}\ee
Then, making use of (\ref{82}), Eq. (\ref{1}) provides the reduced equation
\be \rho^{2}_{t}\,-\,\rho \rho_{tt}\,-\,\ep \rho \rho_{t}\,=\,1,
\lab{83}\ee
which is transformed in the (ordinary) equation of the modified Liouville
type
\be w_{tt}\,+\,\ep w_{t}\,=\,-\ep^{-2w}
\lab{84}\ee
through the change of variable $\rho\,=\,e^{w}$. We notice that by
setting $w\,=\,-\beta$, Eq. (\ref{84}) coincides with Eq. (\ref{54})
for $c_{1}\,=\,2$.

This property may be interpreted as a hint of a possible connection
between the prolongation algebra and the symmetry generators coming from
the group analysis. This important aspect pertinent to the algebraic theory
of nonlinear field equations is a challenging subject for further
investigations.

For $\ep =0$, Eq.
 (\ref{84}) is solved by
\be w\,=\,\ln \frac{\cos \sqrt{c}(t\,-t_{0})}{\sqrt{c}},
\lab{85}\ee
where $c$ and $t_{0}$ are positive constants.

Therefore, from (\ref{82}) we obtain the exact solution
\be u\,=\,2 \ln\frac{\sqrt{c}x}{\cos \sqrt{c}(t-t_{0})}
\lab{86}\ee
to Eq. (\ref{1}) (with $\ep=0$).

We point out that the function (\ref{86}) can be derived from Eqs. (\ref{50})
and (\ref{51}) as a special case. In fact by choosing $\ep = 0$,
Eqs.(\ref{48}) and (\ref{49}) yield

\be y_{3x}\,=\,(\lambda_{2}-\lambda_{1}x) u_{t}, \qquad\qquad
y_{3t}\,=\,e^{u}[\lambda_{1}\,+\,(\lambda_{2}-\lambda_{1}x) u_{x}],
\lab{87}\ee
from which
\be e^{u}\,=\,\frac{1}{2} \gamma_{t} x^{2},
\lab{88}\ee
where we have assumed $u_{t}\,=\,\gamma(t),$ $\lambda_{2}=0,$ and the
functions of integration have been taken equal to zero. From Eq.(\ref{88})
we get $u_{t}\,=\,\frac{\gamma_{tt}}{\gamma_{t}}\,\equiv\,\gamma$, i.e.
$\gamma(t)\,=\,2 \sqrt{c} \tan\sqrt{c}(t-t_{0}).$ Consequently, Eq.(\ref{88})
reproduces just the solution (\ref{86}).

This result is not surprising, since the requirement $u_{t}\,=\,\gamma(t)$
means that we single out a class of solutions to Eq.(1) $(\ep=0)$
corresponding to the symmetry reduction based on the invariant $\rho(t)$
associated with the generator $X^{0}_{4}$. The relation between $\rho(t)$
and $\gamma(t)$ is $-2 \frac{\rho_{t}}{\rho}\,=\,\gamma$ [see (\ref{82})].
Finally, the algebra (\ref{43})-(\ref{46}) with $\ep = 0$ coincides with
that derived from (\ref{153}) for $a_{1}=0$ (see Appendix A).

Second, let us deal with the vector field obtained from the linear
combination of the generators $X_{3}\,=\,X^{0}_{3}=\partial_{x}$ and
$X_{2}\,=\,X^{0}_{2}=\partial_{t}:$
\be V\,=\,v \partial_{x}\,+\,\partial_{t},
\lab{89}\ee
where $v$ is a constant.

The group transformations read $x^{\prime}\,=\,x\,+\,v\lambda$,
$t^{\prime}\,=\,t\,+\,\lambda$, which furnish the invariant
\be \sigma\,=\,x^{\prime}\,-\,v t^{\prime}\,=\,x\,-\,v t.
\lab{90}\ee
This quantity can be exploited to find a self-similar solution to Eq.(1).
In fact, inserting $u\,=\,u(\sigma)$ into Eq.(1) we get
\be (v^{2}\,-\,e^{u})u_{\sigma}\,=\,\ep v u \,+\, c_{0},
\lab{91}\ee
where $c_{0}$ is a constant of integration.

By setting
\begin{equation}
u\,=\,-\tau\,-\,\ln v
\label{92}\end{equation}
and $c_{0}\,=\,-2 \ep v \ln v$, Eq.(\ref{91}) provides
\begin{equation}
\int_{0}^{\tau_{1}}\frac{1\,-\,e^{-\tau{\prime}}}{\tau^{\prime}}\,
d\tau^{\prime}\,=\,
\frac{\epsilon}{v}(\sigma-\sigma_{0})
\label{93}\end{equation}
$(\xi_{0}=$ constant).

The left-hand side of (\ref{93}) is the exponential-integral function
\cite{Abramowitz}
\begin{equation}
{\rm Ein}(\tau)\,=\,{\rm E}_{1}(\tau)\,+\,\ln \tau\,+\,\gamma,
\label{94}\end{equation}
where ${\rm E}_{1}(\tau)\,=\,\int^{\infty}_{\tau}
\frac{e^{-\tau{\prime}}}{\tau^{\prime}}\,d\tau^{\prime}$ and $\gamma$ is
the Euler constant.
%%%%%%%%
Hence, from (\ref{92}), (\ref{93}) and (\ref{94}) we have the
self-similar solution
\begin{equation}
u\,=\,- {\rm Ein}^{-1}\,\left[\frac{\epsilon}{v}(\sigma-\sigma_{0})\right]
\,-\,\frac{c_{0}}{\epsilon v},
\label{95}\end{equation}
where ${\rm Ein}^{-1}(\cdot)$ denotes the inverse function of
(\ref{94}). We notice that the reduced equation (\ref{91}) can be
considered as an ordinary differential equation which defines
the special function ${\rm Ein}^{-1}(\cdot)$.
%%%%%%%%%%%%%%%%%%%%%%%%%%%%%%%%%%%%%%%%%%%%%%%%%%%%%%%%%%%%%%%%%%%%%\

\section{CONCLUSIONS}
We have studied an modified version of a continuous
Toda equation in 1+1 dimensions.

We have applied jointly two procedures: the prolongation method and the
symmetry approach, which are mostly based on the use of algebraic and group
techniques. This strategy reveals efficacious both from a conceptual point
of view and for practical purposes, e.g. for the determination of exact
solutions of the equations under consideration.

For $\ep =0$, on the basis of heuristic arguments
we have found that an infinite-dimensional prolongation Lie
algebra may be associated with Eq. (\ref{1}). This algebra can be closed
step by step in the sense explained in Sec.III. In correspondence of any
representation of a given finite-dimensional Lie algebra in terms of
pseudopotential variables, one obtains a linear problem for Eq.(\ref{1}).
Moreover, a link is established between Eq.(\ref{1})
and the wave equation (\ref{28}). This connection derives from a special
realization of the Lie algebra of the Euclidean group $E_{2}$ related to a
class of solutions of the prolongation equations (\ref{5})-(\ref{7})
($\ep$=0). It is noteworthy that nontrivial solutions to Eq.(\ref{28})
expressed in terms of particular Bessel functions are determined. We remark
also that explicit forms of the pseudopotential can be used to solve
certain second order ordinary differential equations with non constant
coefficients (see, for example, Eq.(\ref{49})).

For the case $\ep \ne 0$, we have shown that the prolongation
algebra is finite-dimensional and is constituted by four elements. A matrix
representation of this algebra (see Eq.(\ref{55})) allows us to write
Eq.(\ref{1}) in a potential form which leads to solutions associated with
those admitted by a modified version of the reduced Liouville equation
(\ref{62}). This equation has the same form of that coming from the
symmetry approach {\it via} the generator $X_{4}\,\equiv\,X_{4}^{0}\,=\,
x\partial_{x}\,+\,2\partial_{u}$. This property indicates the existence
of a possible connection between the prolongation method and the symmetry
approach. This is an important methodological aspect which deserves a wide
investigation. Here we remark only that any approach to this problem should
not ignore the contribution by Harrison and Estabrook \cite{harrison} where
the fundamental concepts of Cartan's theory of systems of partial
differential equations are exploited for obtaining the generators of their
invariance groups (isogroups). Another interesting result achieved in the
framework of the symmetry approach is given by Eq.(\ref{95}), which tells
us that the inverse of the exponential-integral function turns
out to be defined by the reduced differential equation (\ref{91})
corresponding to the generator $V\,=\,v \partial_{x}\,+\,\partial_{t}$.

We shall conclude with a few comments.

First, we have not tackled the problem of the complete integrability of
Eq.(1) (for $\ep =0$). Anyway, as it happens for other nonlinear partial
differential equations of physical interest \cite{}, the existence of an
infinite-dimensional prolongation algebra is necessary for the
integrability property. However, in this regard, we recall that completely
integrable nonlinear field equations admit Kac-Moody prolongation algebras
endowed with a loop structure \cite{}. Therefore, a definitive answer on the
complete integrability of Eq. (1) ($\ep =0$) is strictly connected with a
deep study of its associated algebra (\ref{159}). The situation is
different for $\ep \ne 0$, in which a finite-dimensional prolongation
algebra is found at the beginning. This feature indicates that Eq.(1) for
$\ep \ne 0$ is not completely integrable.

Second, in the context of the approximate symmetry approach, i.e. when
$\ep$ is a
small parameter, a realization of a finite-dimensional Lie algebra, i.e.
(\ref{70})-(\ref{73}), can be
constructed by introducing the auxiliary operators (\ref{69}). This
realization can be expressed in terms of boson annihilation and creation
operators. The algebra (\ref{70})-(\ref{73}), in some sense, seems to
characterize the approximate symmetry of Eq.(1), but its role is not yet
clear. For instance, may algebras of this type arise in the study of the
approximate symmetries of other perturbative systems? Finally, our results
could be useful in the study of a three-dimensional extension of Eq.(1),
using the same theoretical frameworks of this paper. Concerning this point
it should be interesting to see, in analogy with what happens in our case,
whether the equations of Proposition 1 of Ref. [1]:
\be u_{xx}\,+\,u_{yy}\,+\,(e^{u})_{zz}\,=\,0,\ee
\be w_{xx}\,+\,w_{yy}\,+\,(we^{u})_{zz}\,=\,0,\ee
share a property similar to that linking Eqs.(\ref{16}) and (\ref{20}),
through a representation of a Lie algebra of an extended Euclidean group.

\appendix
\section{:~CLOSED LIE ALGEBRAS FROM (\ref{9})-(\ref{15}) WITH $\ep$ =0}
\lab{appB}

Exploiting the Jacobi identity, from (\ref{10}) and (\ref{11}) we obtain
\be [a_{k}, a_{k^{\prime}}]\,=\,-\frac{1}{k^{\prime}}\left\{
[L,[b_{k^{\prime}}, a_{k}]]\,-\, (k+1) [b_{k^{\prime}},b_{k+1}]
\right\},
\lab{b1}\ee
\be [b_{k^{\prime}}, a_{k}]\,=\,-\frac{1}{k}\left\{
[L,[b_{k},b_{k^{\prime}}]]\,-\, k^{\prime} [b_{k},a_{k^{\prime}}]
\right\},
\lab{b2}\ee
\be [b_{k}, b_{k^{\prime}}]\,=\,-\frac{1}{k^{\prime}}\left\{
[L,[a_{k^{\prime}-1}, b_{k}]]\,-\, k [a_{k^{\prime}-1},a_{k}]
\right\}.
\lab{b3}\ee
Equations (\ref{b2}) and (\ref{b3}) provide
\be [b_{0}, a_{k}]\,=\,\frac{1}{k} [L,[b_{0},b_{k}]],
\lab{b4}\ee
\be [b_{0}, b_{k}]\,=\,- \frac{1}{k} [L,[a_{k-1},b_{0}]],
\lab{b5}\ee
which give
\be [b_{0}, a_{k}]\,=\,\frac{1}{(k!)^{2}} [L,[L, \ldots, [L,[L,[b_{0},a_{1}]]
\ldots],\lab{b6}\ee
and
\be [b_{0}, b_{k}]\,=\,\frac{1}{k[(k-1)!]^{2}} [L,[L,[b_{0},a_{1}]]
\ldots]
\lab{b7}\ee
where the operator $L$ on the right hand side of (\ref{b6}) and (\ref{b7})
appears 2(k-1) and 2(k-1)-1 times, respectively.

Other useful relations are
\be [a_{0}, a_{k}]\,=\,-\frac{1}{k}\left\{
[L,[b_{k}, a_{0}]]\,-\, (k+1) [b_{k},b_{1}]
\right\},
\lab{b8}\ee
\be [a_{0}, b_{k}]\,=\,\frac{1}{k}\left\{
[L,[a_{0},a_{k+1}]]\,+\, [a_{k-1},b_{1}]
\right\},
\lab{b9}\ee
\be [a_{k}, b_{k}]\,=\,\frac{1}{k}\left\{
[L,[a_{k},a_{k-1}]]\,+\, (k+1) [a_{k-1},b_{k+1}],
\right\},
\lab{b10}\ee
where
\be [a_{k}, a_{k-1}]\,=\,\frac{1}{k}
[L,[b_{k},a_{k-1}]].
\lab{b11a}\ee
We have also
\be [b_{k-1},b_{k}]\,=\,\frac{1}{k} [L,[b_{k-1},a_{k-1}]].\lab{b11b}\ee
Now, let us suppose that $a_{j}\,=\,0,$ for j\,=\,2,3,... . Then, from
Eq. (\ref{9}), (\ref{10}) and (\ref{11}) we obtain
\be [b_{0}, L]\,=\,0,\; [L,a_{0}]\,=\,b_{1},\;[L,b_{1}]\,=\,a_{1}.
\lab{b11c}\ee
Furthermore, from (\ref{b4}) and (\ref{13}):
\be [b_{0}, a_{1}]\,=\,0,\; [a_{0},b_{1}]\,=\,0,
\lab{b11d}\ee
while
\be [a_{0}, a_{1}]\,=\,0
\lab{b11e}\ee
from (\ref{b8}). We have also
\be [b_{0}, b_{1}]\,=\,0
\lab{b11f}\ee
by commuting Eq.(\ref{12}) with $L$. Finally, (\ref{b10}) provides
\be [a_{1}, b_{1}]\,=\,0
\lab{b11g}\ee
since $b_{2}=0$.
The previous commutation relations define the algebra (\ref{18}).

%%%%%%%%%%%%%%%%%%%%%%%%%%%%%%%%%%%%%%%%%%%%%%%%%%%%%%%%%%%%%%%%
Now, let us suppose that $a_{j}\,=\,0,$  $b_{j}\,=\,0,$ for j\,=\,3,4,... .
Then, Eq. (\ref{7}) yields
\be [a_{0}, b_{0}]\,=\,0,
\lab{b12a}\ee
\be [a_{0}, b_{1}]\,+\,[a_{1},b_{0}]\,=\,0,
\lab{b12b}\ee
\be [a_{0}, b_{2}]\,+\,[a_{1},b_{1}]\,+\,[a_{2},b_{0}]\,=\,0,
\lab{b12c}\ee
\be [a_{1}, b_{2}]\,+\,[a_{2},b_{1}]\,=\,0,
\lab{b12d}\ee
\be [a_{2}, b_{2}]\,=\,0.
\lab{b12e}\ee
By commuting (\ref{b12a}) with $L$ and using the Jacobi identity, with the
help of (\ref{9}) ($\ep$ = 0) and (\ref{11}) we find
\be [b_{0}, b_{1}]\,=\,0.
\lab{b13}\ee
Taking account of (\ref{b13}), Eq. (\ref{b4}) gives
\be [b_{0}, a_{1}]\,=\,0.
\lab{b14}\ee
Hence, from (\ref{b12b}):
\be [a_{0}, b_{1}]\,=\,0.
\lab{b15}\ee
Furthermore, Eqs. (\ref{b11a}), (\ref{b15}) and (\ref{b10}) give
 \be [a_{1},a_{0}]\,=\,0,\lab{b16}\ee
\be [a_{1},b_{1}]\,=\,0.\lab{b17}\ee
Since
\be [a_{2},b_{0}]\,=\,0\lab{b18}\ee
(see (\ref{b6}) and (\ref{b14})), from (\ref{b12b}) we deduce
\be [a_{0},b_{2}]\,=\,0.\lab{b19}\ee
On the other hand, from (\ref{b3}) (see (\ref{b17})):
\be [b_{1},b_{2}]\,=\,0,\lab{b20}\ee
while (\ref{b2}) provides
\be [b_{2},a_{1}]\,=\,2 [b_{1},a_{2}].\lab{b21}\ee
Keeping in mind (\ref{b21}), from (\ref{b12d}) we have
\be [b_{1},a_{2}]\,=\, [b_{2},a_{1}]\,=\,0.\lab{b22}\ee
We get also
\be [a_{2},a_{1}]\,=\,0,\lab{b23}\ee
from (\ref{b1}) and (\ref{b2}). Finally, (\ref{b8}) gives
(see (\ref{b19}) and (\ref{b20})):
\be [a_{0},a_{2}]\,=\,0.\lab{b24}\ee
Thus, all the commutators among the elements $a_{j}$, $b_{j}$ ($j$ = 0,1,2)
have been determined. The commutators of the type $[a_{j},L]$ and $[b_{j},L]
$
are expressed by (\ref{10}) and (\ref{11}). Therefore, we have the seven-
dimensional Lie algebra (\ref{22}).

Another example of finite-dimensional Lie algebra emerging from the
commutation relations (\ref{9})-(\ref{15}), involves the nine generators
$a_{j}$, $b_{j}$ and $L$ with $j=0,1,2,3.$ This algebra can be obtained
following essentially the same scheme adopted for the construction of the
algebra (\ref{22}).

In fact, by putting
$$ M\,=\, a_{0}\,+\,a_{1} z\,+\,a_{2} z^{2}\,+\,a_{3} z^{3},
$$
\be P\,=\, b_{0}\,+\,b_{1} z\,+\,b_{2} z^{2}\,+\,b_{3} z^{3},
\lab{b25}\ee
Eq.(\ref{7}) entails
\be [a_{0}, b_{0}]\,=\,0,
\lab{b26a}\ee
\be [a_{0}, b_{1}]\,+\,[a_{1},b_{0}]\,=\,0,
\lab{b26b}\ee
\be [a_{0}, b_{2}]\,+\,[a_{1},b_{1}]\,+\,[a_{2},b_{0}]\,=\,0,
\lab{b26c}\ee
\be [a_{0}, b_{3}]\,+\,[a_{1}, b_{2}]\,+\,[a_{2},b_{1}]\,+\,[a_{3},b_{0}]
\,=\,0,
\lab{b26d}\ee
\be [a_{1}, b_{3}]\,+\,[a_{2},b_{2}]\,+\,[a_{3},b_{1}]\,=\,0,
\lab{b26e}\ee
\be [a_{2}, b_{3}]\,+\,[a_{3},b_{2}]\,=\,0,
\lab{b26f}\ee
\be [a_{3}, b_{3}]\,=\,0.
\lab{b26g}\ee
Equations (\ref{b26a})-(\ref{b26c}) have been already examined. We need to
scrutinize only those relations which contain $a_{3}$ and $b_{3}$, i.e.
Eqs. (\ref{b26d})-(\ref{b26g}).

\noindent Since
\be [a_{3}, b_{0}]\,=\,0,
\lab{b27}\ee
(see (\ref{b6}) and (\ref{b14})), Eq.(\ref{b26d}) becomes
\be [a_{0}, b_{3}]\,+\,[a_{1}, b_{2}]\,+\,[a_{2},b_{1}]\,=\,0.
\lab{b28}\ee
Now, by commuting (\ref{b24}) and (\ref{b20}), we easily find
\be 3[a_{0}, b_{3}]\,=\,[a_{2}, b_{1}],
\lab{b29}\ee
\be 2[b_{1}, a_{2}]\,=\,[b_{2}, a_{1}].
\lab{b30}\ee
Combining together (\ref{b28}), (\ref{b29}) and (\ref{b30}) we arrive at
\be [a_{0}, b_{3}]\,=\,0,
\lab{b31}\ee
\be [a_{2}, b_{1}]\,=\,0,
\lab{b32}\ee
\be [a_{1}, b_{2}]\,=\,0.
\lab{b33}\ee

{}From (\ref{b8}) and (\ref{b31}):
\be [a_{0},a_{3}]\,=\,\frac{1}{3} [b_{3}, b_{1}].\lab{b34}\ee
On the other hand, Eq. (\ref{b31}) and (\ref{b33}) give
\be [a_{1}, a_{2}]\,=\,0.
\lab{b35}\ee
Thus, using (\ref{b3}), (\ref{b32}) and (\ref{b35}):
\be [b_{1}, b_{3}]\,=\,0.
\lab{b36}\ee
Hence, from (\ref{b34}):
\be [a_{0}, a_{3}]\,=\,0.
\lab{b37}\ee
At this stage, let us consider the relation (\ref{b10}) for $k=2$. We have
\be [b_{2}, a_{2}]\,=\,\frac{3}{2}[a_{1}, b_{3}].
\lab{b38}\ee
Furthermore, from (\ref{b2}) and (\ref{b36}):
\be [b_{3}, a_{1}]\,=\,3[b_{1}, a_{3}].
\lab{b39}\ee
Now, by elaborating the commutator $[L,[a_{0},a_{3}]]$ by means of the
Jacobi identity (see (\ref{b37})) and taking $b_{4}$=0, we get
\be [a_{3}, b_{1}]\,=\,0.
\lab{b40}\ee
Thus, (\ref{b39}) implies
\be [b_{3}, a_{1}]\,=\,0,
\lab{b41}\ee
so that (\ref{b38}) yields
\be [a_{2}, b_{2}]\,=\,0.
\lab{b42}\ee
Inserting (\ref{b42}) in  (\ref{b11b}) for $k=3$, we have
 \be [b_{2}, b_{3}]\,=\,0.
\lab{b43}\ee
Then, from (\ref{b1}):
\be [a_{1}, a_{3}]\,=\,0,
\lab{b44}\ee
{\it via} (\ref{b36}) and (\ref{b43}).

To conclude, Eq.(\ref{b36}) provides
\be 3[b_{2}, a_{3}]\,=\,2[b_{3}, a_{2}]
\lab{b45}\ee
[see (\ref{b43})]. Therefore, Eq. (\ref{b45}) and (\ref{b26f}) furnish
\be [b_{3}, a_{2}]\,=\,0,
\lab{b46}\ee
\be [b_{2}, a_{3}]\,=\,0.
\lab{b47}\ee
We have also
\be [a_{2}, a_{3}]\,=\,0,
\lab{b48}\ee
from (\ref{b1}) and (\ref{b46}).
Therefore, by collecting all the results achieved in the case $a_{j}=0$,
$b_{j}=0$ $(j=4,5,\ldots),$ the finite-dimensional Lie algebra of elements
$a_{0},\,a_{1},\,a_{2}\,a_{3},\,b_{0},\,b_{1},\,b_{2},b_{3}$ and $L$ turns
out to be
defined by the commutation relations:
$$ [a_{j},a_{k}]\,=\,0,\quad [a_{j},b_{k}]\,=\,0,\quad [b_{j},b_{k}]\,=\,0,
$$
$$ [L,a_{l-1}]\,=\,l b_{l},\quad [L,a_{3}]\,=\,0,\; [L,b_{k}]\,=\,k a_{k},
$$
for $j,k=0,1,2,3$ and $l=1,2,3.$

%%%%%%%%%%%%%%%%%%%%%%%%%%%%%%%%%%%%%%%%%%%%%%%%%%%%%%%%%%%%%%%%%%%%%%%%%
\appendix
\setcounter{section}{1}
\section{:~FINITE DIMENSIONAL PROLONGATION ALGEBRA ($\ep \ne  0$)}
\lab{appA}
Here we prove that the prolongation algebra associated  with
Eq. (\ref{1}) with $\ep \ne 0$ is the finite-dimensional Lie algebra
(\ref{43})-(\ref{46}). In doing so, let us start from the commutation rule
\be [L, b_{1}]\,=\,a_{1},
\lab{a1}\ee
coming from (\ref{10}) for $k=1$.

Then, by exploiting the Jacobi identity we have
\be [L,[b_{0},b_{1}]]\,+\,\ep a_{1}\,=\,[b_{0},a_{1}],
\lab{a2}\ee
with the help of (\ref{9}). On the other hand, the relation (\ref{11})
(with $k=1$) provides
\be [b_{0},b_{1}]\,=\,\ep b_{1},
\lab{a3}\ee
by elaborating the commutator $[b_{0},[L,a_{0}]]$ {\it via} the Jacobi
identity. Now, substitution from (\ref{a3}) into (\ref{a2}) gives
\be [b_{0}, a_{1}]\,=\,2\ep a_{1}.
\lab{a4}\ee
Taking account of (\ref{a4}), Eq.(\ref{13}) yields
\be [a_{0},b_{1}] = 2\ep a_{1}.
\lab{a5}\ee
At this point, let us consider the commutator $[b_{1}[a_{0},b_{0}]]$
(see (\ref{12})), By resorting again to the Jacobi identity and using
(\ref{a3}), (\ref{a4}) and (\ref{a5}) we obtain the constraint
\be \ep^{2} a_{1}\,=\,0.
\lab{a6}\ee
Let us suppose that $\ep \ne 0$. Hence, Eq.(\ref{a6}) implies $a_{1}\,
\equiv 0$. Consequently, from Eqs.(\ref{10}) and (\ref{11}) we infer that
$2 b_{2}\,=\,[L,a_{1}]\,=\,0,$ \quad $2 a_{2}\,=\,[L,b_{2}]\,=\,0,$
$3 b_{3}\,=\,[L,a_{2}]\,=\,0,$ \quad $3 a_{3}\,=\,[L,b_{3}]\,=\,0,$  and so on.
Therefore, for $\ep\,\ne 0$ the quantities $a_{0}$, $a_{1}$, $b_{0}$,
$b_{1}$, and $L$, define the closed Lie algebra $\cal L$ expressed by
(\ref{43})-(\ref{46}).
%%%%%%%%%%%%%%%%%%%%%%%%%%%%%%%%%%%%%%%%%%%%%%%%%%%%%%%%%%%%%%%%%%%%
\appendix
\setcounter{section}{2}
\section{:~THE APPROXIMATE LIE GROUP ANALYSIS}
\lab{appC}
Here we recall some basic concepts inherent to the approximate group
analysis of differential equations derived by Baikov, Gazizov and Ibragimov
(BGI) in 1988 \cite{Ibragimov}. The starting point of this approach is a
Theorem (see below) that allows one to construct approximate symmetries
which are stable for small perturbations of the differential equation under
investigation.

To describe briefly the BGI method let us consider the one-parameter group
of local point transformations:
\be z^{\prime}\,=\,g(z,\ep ,a),
\lab{c1}\ee
where $z=(z_{1},\ldots,z_{N})$ is the independent variable, and $a$ is the
group parameter so that the value $a=0$ corresponds to the identity
transformation $g(z,\ep,a)=z,$
\be g(g(z,\ep ,a),\ep,b)\,=\,g(z,\ep ,a+b),
\lab{c2}\ee
and $\ep $ is a perturbative parameter. Then let us suppose that
$f \approx g$, namely
\be f(z,\ep,a)\,=\,g(z,\ep ,a)\,+\,o(\ep^{p})
\lab{c3}\ee
for some fixed values of $p \ge 0.$

The transformations
\be z^{\prime}\, \approx\,f(z,\ep,a),
\lab{c4}\ee
form an approximate one-parameter group $f$
\be f(z,\ep,0) \,\approx\, z ,\lab{c5}\ee
\be f(f(z,\ep,a), \ep, b)  \,\approx\,f(z, \ep, a+b) ,\lab{c6}\ee
and the condition $f(z,\ep, a)\, \approx\,z$ for all $z$ implies that
$a$ =0.

The following Theorem holds:
\begin{teo}
Let us assume that the transformations (\ref{c4}) form an approximate group
with the tangent vector field
\be \xi(z,\ep) \,\approx\,\frac{\partial f(z, \ep, a)}{\partial a}|_{a=0}.
\lab{c7}\ee
Then, the function $f(z,\ep ,a)$ satisfies
\be \frac{\partial f(z,\ep,a),}{\partial a}  \,\approx\, \xi(f(z, \ep, a)).
\lab{c8}\ee
Conversely, for any (smooth) function $\xi(z,\ep)$ the solution (\ref{c4})
of the approximate Cauchy problem
$$\frac{d z^{\prime}}{d a}  \,\approx\,\xi(z^{\prime}, \ep), $$
\be z^{\prime}|_{a=0} \,\approx\,z , \lab{c9}\ee
determines an approximate one-parameter group with parameter $a$.
\end{teo}

The previous Theorem is called the approximate Lie theorem, while Eq.
(\ref{c9}) is called the approximate Lie equation.

Resorting to the approximate Lie Theorem, we can construct the approximate
group of transformations generated by a given infinitesimal operator. To
see how the method works out, let us deal with the case $p=1$. Then we look
for the approximate group of transformations
\be z^{\prime} \,\approx\,f_{0}(z ,a)\,+\,\ep f_{1}(z,a), \lab{c10}\ee
determined by the infinitesimal operator
\be X \,=\,\left(\xi_{0}(z)\,+\,\ep \xi_{1}(z)\right)
\frac{\partial}{\partial z}. \lab{c11}\ee
The related approximate Lie equation
\be\frac{d}{da}(f_{0}\,+\,\ep f_{1}) \,\approx\,\xi_{0}
(f_{0}(z)\,+\,\ep f_{1}) \,+\,\ep \xi_{1}(f_{0}(z)\,+\,\ep f_{1})
 \lab{c12}\ee
becomes the system
\be\frac{d f_{0}}{da}\,\approx\,\xi_{0}
(f_{0}), \quad\frac{d f_{1}}{da}\,\approx\,\xi^{\prime}_{0}
(f_{0})f_{1}\,+\, \xi_{1}(f_{0}),
 \lab{c13}\ee
where $\xi^{\prime}_{0}$ is the derivative of $\xi_{0}.$

\noindent The initial condition $z^{\prime}|_{a=0}\,\approx\,z$ provides
 $f_{0}|_{a=0}\,\approx\,z$ and $f_{1}|_{a=0}\,\approx\,0.$

Now, we are ready to introduce the concept of approximate invariance.

Precisely, the approximate equation
\be F(z,\ep)\,\approx\,0
\label{c14}\ee
is said to be invariant with respect to the approximate group of
transformations
$$z^{\prime}\,\approx\,f_{0}(z,\epsilon,a)$$
if
\begin{equation}
F(f(z,\epsilon,a))\,\approx 0 \label{c15}
\end{equation}
 for all $z$ satisfying (\ref{c15}).
A criterion for obtaining the approximate symmetries of a given equation is
expressed by
\begin{teo}

Let us suppose that the function $F(z,\epsilon)\,=\,\left(F^{1}(z,
\epsilon),\ldots,F^{n}(z,\epsilon)\right)$, $n < N$, which is jointly
 analytic in the variables $z$ and $\epsilon$, satisfies the condition
$$ {\rm rank}\, F^{\prime}(z,0)\mid_{F(z,0)=0}\,=\,n,$$
where
$F^{\prime}(z,0)\,=\,\|\frac{\partial F^{\nu}(z,0)}{\partial z^{i}}\|$ for
$\nu=1,\dots,n$ and $i=1,\dots,N$. For the equation
\begin{equation}
F(z,0)\,=\,o(\epsilon^{p})\label{c16}
\end{equation}
to be invariant under the approximate group of transformations
\begin{equation}
z^{\prime}\,=\,f(z,\epsilon,a)\,+\,o(\epsilon^{p})\label{c17}
\end{equation}
with infinitesimal operator
\begin{equation}
X\,=\,\xi(z,\epsilon)\,\frac{\partial}{\partial z},\qquad\qquad
\xi\,=\,\frac{\partial f}{\partial a}|_{a=0}\;\;+\;\;o(\epsilon^{p})
\label{c18}\end{equation}
it is necessary and sufficient that
\begin{equation}
XF(z,\epsilon)\,=\,o(\epsilon^{p})\label{c19}
\end{equation}
whenever $F(z,\epsilon)\,=\,o(\epsilon^{p})$.
\end{teo}

In order to solve Eq.(\ref{c18}) to within $o(\epsilon^{p})$
one needs to represent $z$, $F$ and $\xi^{k}$ in the form
$$ z\,\approx\,y_{0}\,+\,\epsilon y_{1}\,+\,\ldots\,+\,\epsilon^{p} y_{p},
\qquad F(z,\epsilon)\,\approx\,\sum^{p}_{i=0}\,\epsilon^{i}F_{i}(z),$$
$$ \xi^{k}(z,\epsilon)\,\approx\,\sum^{p}_{i=0}\,\epsilon^{i}\xi^{k}_{i}
(z).$$
For $p=1$, we obtain
\begin{equation}
\xi_{0}^{k}(y_{0})\,\frac{\partial F_{0}}{\partial z^{k}}|_{z=y_{0}}\,=\,0,
\label{c20}\end{equation}

\begin{equation}
\xi_{1}^{k}(y_{0})\,\frac{\partial F_{0}}{\partial z^{k}}|_{z=y_{0}}\,+\,
\xi_{0}^{k}(y_{0})\,\frac{\partial F_{1}}{\partial z^{k}}|_{z=y_{0}}\,+\,
y^{l}_{1}
\,\frac{\partial}{\partial z^{l}}\left[
\xi_{0}^{k}(z)\,\frac{\partial F_{0}}{\partial z^{k}}\right]|_{z=y_{0}}\,=\,0
\label{c21}\end{equation}
under the conditions
$$F_{0}(y_{0})\,=\,0,\qquad\qquad F_{1}(y_{0})\,+\,y^{l}_{1}\,
\frac{\partial F_{0}(z)}{\partial z^{l}}|_{z=y_{0}}\,=\,0.$$

%%%%%%%%
%%%%%%%%%%%%%%%%%%%%%%%%%%%%%%%%%%%%%%%%%%%%%%%%%%%%%%%%%%%%%%%%%%%

%%%%%%%%
%%%%%%%%%%%%%%%%%%%%%%%%%%%%%%%%%%%%%%%%%%%%%%%%%%%%%%%%%

\end{document}